\documentstyle[12pt]{article}
\title{The ideas of gravitational effective field theory}
\author{John F. Donoghue\\[5mm]
Department of Physics and Astronomy\\
University of
Amherst, MA ~01003}
\date{}
\begin{document}
\begin{titlepage}
\maketitle
\begin{abstract}

I give a very
brief introduction to the use of effective field theory techniques
in quantum
calculations of general
relativity.  The gravitational interaction is naturally
organized as a quantum
effective field theory and a certain class of  quantum corrections can be
calculated. [Talk presented at the XXVIII
International Conference on High Energy
Physics, ICHEP94, Glasgow, Aug. 1994, to be published in the proceedings.]

\end{abstract}
{\vfill UMHEP-413 \\ hep-th/9409143}
\end{titlepage}

We expect that there will be new interactions and new degrees of freedom at
the Planck scale, if not
sooner.  Many discussions of quantum mechanics and gravity involve
speculations about Planck
scale physics.  This talk describes a more conservative approach as we will
use quantum
mechanics and general relativity at ordinary energies, where we expect that
they should both be
valid.  The goal is to argue that general relativity forms a fine quantum
theory at ordinary energies,
and to identify a class of ``leading quantum corrections'' which are the
dominant quantum effects at
long distance and which are reliably calculable.  The apparent obstacle to
such a program is the fact
that the quantum corrections involve integration over all energy scales,
including extreme high
energies.  The solution to this is the use of effective field theory.
Because of the briefness of this report, the
discussion here is
necessarily superficial, but I will concentrate on the basic ideas of such an
approach [1].

Effective field theory is a technique
that allows one to separate the effects of
high energy scales
from low energy ones.  In many cases, such as the theory of gravity, one
does not know the
correct high energy theory.  However as a consequence of the uncertainty
principle we do know
that,  when viewed at low energy, the high energy degrees of freedom do not
propagate far.  They
can be integrated out of the theory leaving a local Lagrangian, although this
Lagrangian will in
general contain nonrenormalizable interactions.  In contrast, the low energy
degrees of freedom
propagate long distances and cannot be summarized by a local interaction.
They must be included
explicitly.  From an unknown high energy theory, we are then led to write
the most general
Lagrangian containing the low energy particles which is consistent with the
symmetries and
vacuum structure of the theory.  In the case of gravity interacting with a
massive matter field we
impose general covariance and find that

\begin{eqnarray}
{\cal L} & = & {\cal L}_{gr} + {\cal L}_{matter} \nonumber \\
{\cal L}_{gr} & = &\sqrt{-g}
\left\{ \Lambda + {2 \over \kappa^2} R + c_1 R^2 +
c_2 R_{\mu \nu} R^{\mu \nu} + \dots \right\} \nonumber \\
{\cal L}_{matter} & = &\sqrt{-g} \left\{ {1 \over 2}
\left( g^{\mu \nu} \partial_{\mu} \phi
\partial_{\nu} \phi - m^2 \phi^2 \right) \right. \nonumber \\
& + & \left. d_1 R_{\mu \nu} \partial^{\mu} \phi \partial^{\nu}
\phi \right.\nonumber \\
 & + & \left. R \left( d_2 \partial_{\lambda} \phi
\partial^{\lambda} \phi - d_3 m^2 \phi^2
\right)
+\ldots \right\}
\end{eqnarray}

\noindent where $\Lambda \approx 0$ is related to the cosmological
constant (we will set this equal
to zero), $\kappa^2 = 32 \pi G$, and $c_i , d_i$ are unknown constants.  The
second key ingredient to
effective field theory is the energy expansion, in which the many terms in
the effective Lagrangian
are ordered in powers of the low energy scale over the high energy
scale.  In gravity, since
$R$ involves two
derivatives (which will become two factors of momentum $q$ in matrix
elements), the $R^2$
terms will be of order $q^4$ and hence much smaller than the $R$ term at low
enough energies.  It is for
this reason that we have essentially no phenomenological constraint on the
$R^2$ terms $(i.e. c_1, c_2
< 10^{74})$ [2].

While it may seem relatively obvious that a classical Lagrangian can be
ordered in an energy
expansion, it is perhaps less obvious that quantum effects of the low energy
particles can also be
so ordered[3].  However, in loop diagrams the  high momentum portions of the
integration and all the ultraviolet divergences are also equivalent to
local counterterms in a
Lagrangian.  For example the effect of gravitons at one-loop order
has the high
energy behavior (in
dimensional regularization) of [4]

\begin{equation}
{\cal L}_{div} = \sqrt{-g} {1 \over 8\pi^2(4-d)} \left({1 \over 120}R^2
+ {7\over 20}R_{\mu \nu} R^{\mu
\nu} \right).
\end{equation}

\noindent This is probably not an accurate description of the full high energy
behavior, but this
does not matter because such quantum effects are not themselves observable
and can be absorbed into
renormalized values of the
unknown constants $c_i$.  However within the same Feynman diagrams
there are also low energy quantum
effects which correspond
to the long range propagation of gravitons.  These are reliable because they
are independent of the unknown high energy theory, depending only on the
massless degrees of freedom (gravitons) and their couplings at the lowest
energies (which follow from the Einstein action). In calculations, the
distinguishing characteristic is the analytic structure of the amplitudes.
Effects which are able to be expanded in a power series in the momentum are
thereby in a form that has the same structure as operators which arise
in a local Lagrangian. These are then in most cases indistinguishable from
possible effects from a high energy theory, which we argued above would
be contained in the unknown coefficients of a local Lagrangian. However,
non-analytic effects in the matrix element cannot come from a local
Lagrangian, and only arise from long range propagation of light particles.
One can use this distinction to separate out the low energy quantum effects. In
most cases, the nonanalytic terms are larger numerically when one works at
extremely low energies, so they are the leading long distance corrections.
Effective field theory is a procedure which carries out these ideas in
a straightforward way.

These ideas can perhaps be best explained by an example.  The usual
gravitational interactions
between two masses can be obtained from the one graviton exchange
potential

\begin{equation}
{\kappa^2 m_1 m_2 \over 8 q^2} \rightarrow - {G m_1 m_2 \over r}
\end{equation}

\noindent The effects of the $R^2$ terms in the effective Lagrangian appear
at one higher power of
$q^2$.  Loop diagrams also give contribution at this order but the nonlocal
effects of low energy
are represented by nonanalytic terms in momentum space, i.e.,
schematically

\begin{eqnarray}
V(q) \sim \kappa^2 m_1 m_2 \left[ {1 \over q^2}
+ \left( c_i + \kappa^2 \ell_i \right) \right.
\nonumber \\
\left. + \kappa^2 \left( a \sqrt{{m^2 \over -q^2}}
+ b ln (-q^2) \right) + \ldots
\right]
\end{eqnarray}

\noindent where $a, b $ and $\ell_i$ arise form the calculation of a set of
one-loop diagrams. ($\ell_i$ is divergent and is absorbed into the renormalized
value of the parameter $c_i$.)
The coefficient of the nonanalytic terms ($a, b$) are finite and are a
consequence of the low
energy part of the theory.  While most work in the field has focussed on the
divergent portion, it is these latter finite terms which are the most
predictive part of the diagrams. Note that at low enough momentum,
the non-analytic terms are larger than the constant terms. In addition,
they are distinguished by a different spatial dependence. When one forms a
nonrelativistic potential one
finds

\begin{eqnarray}
V(r) &\sim& -{Gm_1 m_2 } \left[ {1 \over r} + 4\pi\left( c_i + \kappa^2
\ell_i \right) \delta^3 (x) \right.
\nonumber \\
 &+&\left.   {2 \over \pi}
{\kappa^2 am \over r^2} -{2 \over\pi}{\kappa^2 b\over  r^3} + \ldots \right]
\end{eqnarray}

\noindent The nonanalytic terms give power-law corrections while the local
Lagrangian and high energy loop effects give a delta function. Thus the long
distance quantum correction to the Newtonian potential is calculable.  An
explicit calculation [1]
yields

\begin{eqnarray}
V(r) = - {Gm_1 m_2 \over r} \left[ 1 - {G(m_1 + m_2) \over r c^2} \right.
\nonumber \\
\left. - {127 G  \hbar \over 30 \pi^2 r^2 c^3} + \ldots \right]
\end{eqnarray}

The idea of a gravitational effective field theory extends well beyond this
calculation.  In general,
effective field theory techniques will organize any given matrix element into
the calculable effects
of low energy and the unknown effects of the full high energy theory.  Most
commonly, the
nonanalytic terms are the leading contributions at large distance.  This
division has not been
commonly applied to the gravitational interactions and much of the standard
wisdom of general
relativity needs to be scrutinized through the eyes of effective field theory.
The quantum corrections are numerically small in macroscopic phenomena, and
I know of no such effects that can influence present day experimental
relativity. However, these ideas may be useful in elucidating some of the
theoretical issues of general relativity, and perhaps can be compared to
the work being done in lattice simulations of quantum gravity.

As far as quantum
mechanics are concerned, effective field theories are as natural as the more
restrictive class of
renormalizable field theories (and are perhaps even more natural).  From
this point of view, the
quantum theory of gravity does not seem more problematic at ordinary
energies than the rest of the
Standard Model.

\vspace{5mm}
\noindent
\end{document}